\documentclass[10pt,journal,compsoc]{IEEEtran}

\usepackage{booktabs}
\usepackage{cite}
\usepackage{graphicx}
\usepackage{listings}
\usepackage{xcolor}
\usepackage{url}
\usepackage{balance}
\usepackage{tabularx}
\newcolumntype{Y}{>{\raggedright\arraybackslash}X}
\usepackage{tikz}
\usetikzlibrary{arrows.meta,positioning,calc}

\definecolor{codeblue}{RGB}{32,78,121}
\definecolor{codegray}{RGB}{90,90,90}
\title{Memory as a Temporal Resource:\\
Lifetime Contracts for Multi-Retention Systems}

\author{Adrian Jackson%
\thanks{Adrian Jackson works at EPCC, The University of Edinburgh, Edinburgh, UK.
E-mail: a.jackson@epcc.ed.ac.uk.}%
}
\begin{document}
\maketitle

\begin{abstract}
Computers move and rewrite data whose useful lifetimes range from inner-loop iterations to years, yet memory systems usually hide retention as a fixed physical property. We argue that required lifetime should become a first-class resource across algorithms, programming systems, controllers, and devices. Building on prior lifetime-aware placement, we propose an event-based validity contract that links application completion events to retention-class selection and renewal of protection. Applications would be enabled to express semantic contracts such as ``until this tile completes'' and the system will map data to homogeneous arrays engineered for ephemeral, epoch, or durable retention and preserve correctness through margins, ECC, refresh, migration, or recomputation. We present a trace-based simulator and FPGA prototype that exercises class migration.
\end{abstract}


\section{The Missing Dimension in the Memory Hierarchy}

\IEEEPARstart{A}{I}, and large scale computing, have made memory movement a first-order constraint on performance and energy. Modern accelerators respond with wider off-chip interfaces, large on-chip SRAMs, aggressive data reuse, and recomputation. These responses are necessary, but they optimise where data reside and how quickly data move. They rarely ask a more elementary question: \emph{for how long must each value remain correct?}

That omission creates a mismatch. A matrix-multiplication microtile may be dead after hundreds or thousands of cycles. An activation checkpoint may be required until the end of a training step. Model parameters and recovery metadata may need much longer protection. Nevertheless, the memory system often treats these objects as though they require the same retention. SRAM retains state for as long as power is supplied; DRAM and HBM are refreshed without reference to application semantics; and non-volatile memories are usually designed for lifetimes far beyond those of transient tensors or data structures.

The energy cost of moving and storing data is already material relative to arithmetic~\cite{horowitz2014}. We therefore propose a new abstraction: \emph{memory as a temporal resource}. Just as software states how much memory it needs and hardware decides where to place it, software should be able to state a correctness lifetime while hardware decides how strongly to retain the data. Storage effort can then be proportional to useful lifetimes rather than to the longest-lived object in the machine.

Building on retention-relaxed caches, lifetime-aware scratchpads, and software-managed heterogeneous memories, this Perspective proposes an event-based validity contract for application regions and tensors. Software identifies the semantic event until which contents are needed, the memory system selects a retention class and must renew protection while the event remains incomplete. The contribution is a common contract connecting dataflow boundaries to placement and maintenance obligations, together with a trace-based simulator and FPGA migration prototype for exploring associated costs. Tiled matrix multiplication illustrates the mapping using an established transformation. The reported energy and latency reductions are conditional model projections, while the FPGA results measure migration between logical memory classes.

\section{Retention Is Physical; Lifetime Is Semantic}

Spin-transfer-torque magnetic RAM (STT-MRAM) is an instructive storage substrate.
Its magnetic tunnel junctions offer density and near-zero standby leakage, but the write behaviour and retention depend on physical design and operating conditions. Thermal stability is linked to retention, while a lower energy barrier can reduce switching effort at the cost of a shorter and more variable retention interval~\cite{chun2013}.

In conventional STT-MRAM, retention is principally engineered into the cell; it is not independently chosen for every successful write by simply shortening the write pulse. Pulse duration and amplitude affect switching probability. Once the pulse is removed, a successfully written conventional cell returns to the energy landscape established by its physical construction.

Our baseline architecture proposal consequently uses \emph{homogeneous arrays or tiles} of STT-MRAM style memory, each engineered for a specific retention class. Table~\ref{tab:classes} shows the semantic intent of these classes. The exact times and device parameters associated with these classes must be determined by fabrication, temperature, variation, and the required error probability of the memory ultimately used to construct them.

\begin{table}[t]
\caption{Semantic retention classes. Durations are implementation dependent.}
\label{tab:classes}
\centering
\footnotesize
\begin{tabular}{p{0.17\columnwidth}p{0.31\columnwidth}p{0.39\columnwidth}}
\toprule
Class & Typical role & System obligation \\
\midrule
Ephemeral & Tiles, panels, queues, transient activations & Preserve until a
short phase ends; discard or recompute afterward \\
Epoch & Kernel, iteration, or transaction state & Preserve across a bounded
epoch; refresh or promote if the epoch overruns \\
Durable & Inputs, checkpoints, model state, metadata, operating system & Preserve across long
intervals or power-management events \\
\bottomrule
\end{tabular}
\end{table}

Separate retention classes simplify characterisation, sensing, error correction, and timing. They also make the physical contract testable: a controller chooses a bank with known retention statistics. More speculative devices may one day offer electrically reconfigurable energy barriers for more precise control of the retention of individual data structures, but our cross-layer architecture does not depend on them.

\subsection{Related work}

Reduced retention is an established way to trade STT-RAM persistence for write cost. Sun et al. combined multiple retention levels with refresh and migration\cite{sun2011}, while LARS adapts cache retention to application behaviour\cite{kuan2018}, \cite{kuan2019}. These studies provide architectural foundations for the retention classes used here.

Software-directed retention also predates this proposal. Swaminathan et al. exploited short-lived Java objects and generational garbage collection in a low-retention scratchpad\cite{swaminathan2012}. Rodríguez et al. proposed a multi-retention scratchpad with compiler-directed allocation, selecting retention using loop execution bounds and providing a long-retention fallback\cite{rodriguez2014}. Li et al. rearranged program data layout to reduce refreshes in volatile STT-RAM caches\cite{li2015}. HaVOC combines programmer and compiler information, including lifespan, with runtime placement in heterogeneous scratchpad and non-volatile memories\cite{bathen2012}. Thus object-level placement, compiler involvement, and lifetime annotations are established techniques.

Related DRAM research clarifies two different uses of retention information. RAIDR varies refresh according to measured cell retention\cite{liu2012}. Flikker uses application annotations to distinguish critical and noncritical data, reducing refresh while accepting corruption in the latter\cite{liu2011}. Our required contracts instead aim to preserve live contents to a specified reliability target until semantic completion, approximate contracts must be requested explicitly. Region-based memory management provides a complementary precedent for expressing allocation and deallocation boundaries\cite{tofte1994}. A retention contract additionally addresses or considers the validity of stored contents before those boundaries.

Algorithmic lifetime boundaries are likewise not new. FlashAttention uses tiled dataflow to avoid materializing the full attention matrix\cite{dao2022}, and PagedAttention provides mechanisms for managing and sharing key/value storage\cite{kwon2023}. These are possible sources of completion and reuse events for retention management. The matrix-multiplication example discussed in Section 4 uses established tiling to illustrate how temporary buffers and accumulators can be assigned different validity obligations.

This Perspective synthesizes these lines of work into a proposed event-based validity contract spanning application dataflow, runtime placement, and memory maintenance. Its intended extension is to retain the obligation while a semantic event remains incomplete, allowing renewal when execution overruns a prediction. The trace framework and FPGA migration measurements support exploration of that contract's costs rather than validating novel storage hardware.

\section{The Executable Lifetime Contract}

Building on established lifetime-aware placement approaches, we propose an event-based validity contract for application regions and data structures. Software names the semantic event until which their contents are required, and the contract connects that event to compiler analysis, runtime placement, and memory controller functionality. Its intended extension is to keep the required validity obligation active until the event and all registered consumers have completed, including when execution exceeds its predicted duration, without depending on accurate execution-time prediction. The system selects a suitable retention class and renews protection before expiry whenever the contents remain live. This separates the application’s validity requirement from the physical placement policy, allowing the system to choose refresh, promotion, or other preservation mechanisms while maintaining the contract. For example:

\begin{lstlisting}[language=C++,caption={Illustrative lifetime-aware interface.},label={lst:api}]
auto panel = allocate(bytes,
 retention::until_phase_end,
 guarantee::required);

auto state = allocate(bytes,
 retention::until_epoch_end,
 guarantee::required);

phase_begin("attention-tile");
kernel(panel, state);
phase_end(); // panel is now dead
checkpoint(state); // promote if required
\end{lstlisting}

A required contract associates a region with an explicit completion event and a target reliability level. While that event is incomplete, each live value must remain protected within the characterized operating envelope. If a retention window approaches expiry, the system must complete refresh, promotion, or another valid recovery action before protection lapses. A placement preference may be overridden to meet this obligation. The completion event retires the contents only after all registered consumers, including asynchronous transfers, have completed. Unknown lifetimes use a stronger class. Approximate validity requires a separate, explicit error budget.

For example, a tile may be expected to finish within its initial retention window, but contention may delay one of its consumers. The completion event would then remain pending. Under the proposed contract, the system would be required to renew protection before expiry and preserve the contents until all registered consumers have completed. This example specifies the intended enforcement behaviour.

With this approach, the programmer does not request ``4096 cycles of STT-MRAM.'' Instead, the program names a semantic boundary. A compiler or profiler may infer the same information, and an expert can provide a hint when inference is uncertain. The runtime translates the contract into placement and policy using temperature- and variation-aware device profiles.

Contracts must also distinguish correctness from preference. A \texttt{required} contract obliges the system to maintain data validity. A \texttt{preferred} contract permits fallback to a stronger class. An \texttt{approximate} contract may permit a stated error budget, but approximate
behaviour must never be enabled implicitly. Expiration is not deallocation: it means that content is no longer guaranteed. Ownership and lifetime remain separate language-level concepts in our approach.

At runtime, the controller would have five basic choices before a guarantee is exhausted:

\begin{enumerate}
 \item discard data proven dead;
 \item refresh it in the current class;
 \item promote it to a stronger class;
 \item replicate it with suitable error correction; or
 \item recompute it if recomputation is cheaper than preservation.
\end{enumerate}

\begin{figure*}[t]
\centering

\resizebox{0.98\textwidth}{!}{%
\begin{tikzpicture}[
 node distance=7mm,
 flowbox/.style={
 draw=#1!75!black,
 fill=#1!9,
 rounded corners=2pt,
 very thick,
 minimum height=15mm,
 inner sep=3pt,
 align=center,
 font=\sffamily\small
 },
 stage/.style={
 font=\sffamily\scriptsize\bfseries,
 text=black!65
 },
 flow/.style={
 -{Latex[length=2.5mm]},
 very thick,
 draw=black!70
 },
 monitor/.style={
 -{Latex[length=2.2mm]},
 dashed,
 thick,
 draw=black!55
 },
 feedback/.style={
 -{Latex[length=2.5mm]},
 dashed,
 very thick,
 draw=blue!65!black
 }
]

\node[
 flowbox=blue,
 text width=2.1cm
] (algorithm) {
 \textbf{Algorithm}\\
 and dataflow
};

\node[
 flowbox=cyan,
 text width=2.7cm,
 right=of algorithm
] (contract) {
 \textbf{Lifetime contract}\\[1mm]
 Region or tensor\\
 semantic boundary
};

\node[
 flowbox=violet,
 text width=2.8cm,
 right=of contract
] (policy) {
 \textbf{Compiler and runtime}\\[1mm]
 Placement and\\
 fallback policy
};

\node[
 flowbox=orange,
 text width=3.2cm,
 right=of policy
] (controller) {
 \textbf{Memory controller}\\[1mm]
 Place \(\bullet\) Refresh \(\bullet\) Promote\\
 ECC \(\bullet\) Recompute
};

\node[
 flowbox=green,
 text width=3.6cm,
 right=of controller
] (memory) {
 \textbf{Retention-class arrays}\\[1.5mm]
 {\scriptsize
 Ephemeral \(\quad\) Epoch \(\quad\) Durable}
};

\node[stage, above=2mm of algorithm] {APPLICATION};
\node[stage, above=2mm of contract] {SEMANTIC INTERFACE};
\node[stage, above=2mm of policy] {SYSTEM POLICY};
\node[stage, above=2mm of controller] {CONTROL PLANE};
\node[stage, above=2mm of memory] {PHYSICAL STORAGE};

\draw[flow] (algorithm.east) -- (contract.west);
\draw[flow] (contract.east) -- (policy.west);
\draw[flow] (policy.east) -- (controller.west);
\draw[flow] (controller.east) -- (memory.west);

\node[
 draw=black!55,
 fill=black!4,
 rounded corners=2pt,
 thick,
 minimum width=11.5cm,
 minimum height=11mm,
 inner sep=4pt,
 align=center,
 font=\sffamily\footnotesize,
 below=14mm of controller
] (telemetry) {
 \textbf{Telemetry and policy feedback}\\[1mm]
 {\scriptsize
 Temperature \(\bullet\)
 corrected errors \(\bullet\)
 queueing \(\bullet\)
 capacity pressure \(\bullet\)
 observed lifetimes}
};

\draw[monitor]
 (controller.south) --
 (controller.south |- telemetry.north);

\draw[monitor]
 (memory.south) --
 (memory.south |- telemetry.north);

\draw[feedback]
 (telemetry.north west)
 to[out=165,in=-90,looseness=1.15]
 (policy.south);

\end{tikzpicture}%
}

\caption{Lifetime contracts connect application semantics to physical
retention. Software states how long an object must remain correct; the compiler,
runtime, and controller select and maintain an appropriate retention class.
Measured operating conditions feed back into placement and maintenance policy.}
\label{fig:contract}
\end{figure*}
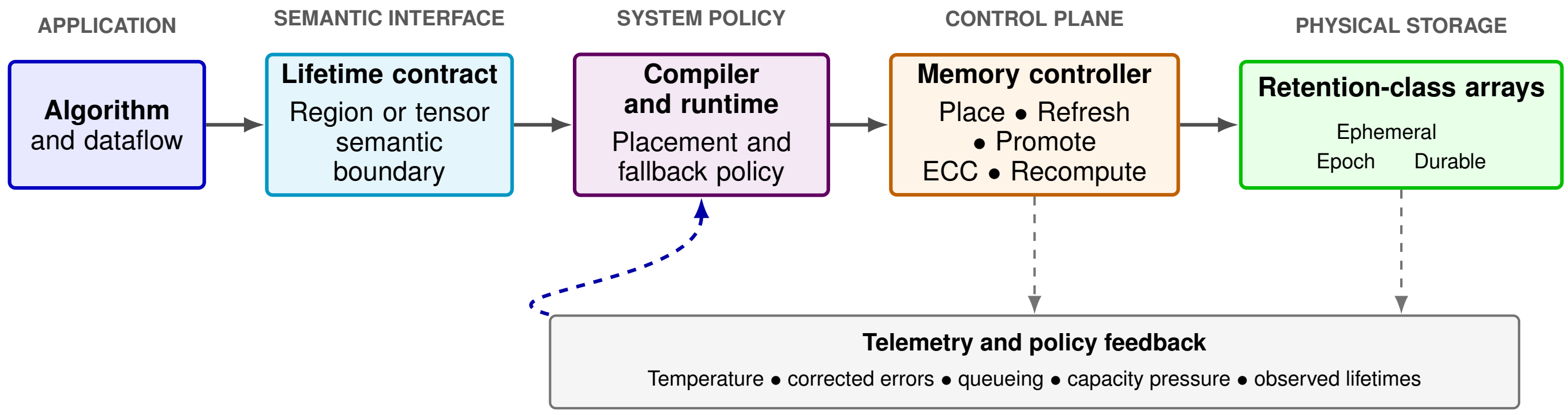

Figure~\ref{fig:contract} summarises the proposed chain, which illustrates the proposed relationship between a semantic lifetime and a measurable hardware obligation.

\section{Algorithms Can Manufacture Short Lifetimes}

Lifetime is not merely a property to track and profile, program transformations can create it. Consider matrix multiplication $C=AB$. In a naive implementation, the input arrays and output remain live for the entire kernel, so a short retention class for those data structures is unsafe even when many elements are accessed close together.

\begin{figure}[t]
\centering

\begin{tikzpicture}[
 node distance=4.5mm,
 data/.style={
 draw=#1!75!black,
 fill=#1!9,
 rounded corners=2pt,
 thick,
 text width=6.2cm,
 minimum height=10mm,
 inner sep=3pt,
 align=center,
 font=\sffamily\footnotesize
 },
 operation/.style={
 draw=orange!80!black,
 fill=orange!9,
 rounded corners=2pt,
 thick,
 text width=5.4cm,
 minimum height=8mm,
 inner sep=3pt,
 align=center,
 font=\sffamily\footnotesize
 },
 flow/.style={
 -{Latex[length=2.2mm]},
 thick,
 draw=black!70
 },
 loop/.style={
 -{Latex[length=2.2mm]},
 dashed,
 thick,
 draw=blue!65!black
 },
 annotation/.style={
 font=\sffamily\scriptsize,
 text=black!65
 }
]

\node[data=green] (inputs) {
 \textbf{Durable inputs}\\
 Complete matrices \(A\) and \(B\)\\[-1mm]
 {\scriptsize Retained for the complete kernel}
};

\node[operation, below=of inputs] (load) {
 Load the next \(k\)-block
};

\node[data=cyan, below=of load] (microtiles) {
 \textbf{Ephemeral working set}\\
 Packed \(A_{i,k}\) microtile
 \(\qquad\)
 Packed \(B_{k,j}\) microtile\\[-1mm]
 {\scriptsize Overwritten when the next \(k\)-block is loaded}
};

\node[operation, below=of microtiles] (mac) {
 Multiply--accumulate\\[-1mm]
 \[
 C_{i,j} \leftarrow
 C_{i,j} + A_{i,k}B_{k,j}
 \]
};

\node[data=violet, below=of mac] (accumulator) {
 \textbf{Epoch state}\\
 Partial \(C_{i,j}\) accumulator tile\\[-1mm]
 {\scriptsize Retained across all \(k\)-blocks for this output tile}
};

\node[data=green, below=9mm of accumulator] (output) {
 \textbf{Durable output}\\
 Committed \(C_{i,j}\) tile
};

\draw[flow] (inputs) -- (load);
\draw[flow] (load) -- (microtiles);
\draw[flow] (microtiles) -- (mac);
\draw[flow] (mac) -- (accumulator);

\draw[flow]
 (accumulator) --
 node[right, annotation] {all \(k\)-blocks complete}
 (output);

\coordinate (loopcorner) at ($(accumulator.east)+(7mm,0)$);

\draw[loop]
 (accumulator.east) --
 (loopcorner) |-
 node[pos=0.60, right, annotation, text=blue!65!black]
 {next \(k\)-block}
 (load.east);

\end{tikzpicture}

\caption{Dataflow and lifetime boundaries for one output tile of blocked matrix
multiplication. The packed input microtiles expire after each inner-loop step,
whereas the partial output tile remains valid across the complete \(k\)-loop
before being committed to durable storage.}
\label{fig:tiled-dataflow}
\end{figure}
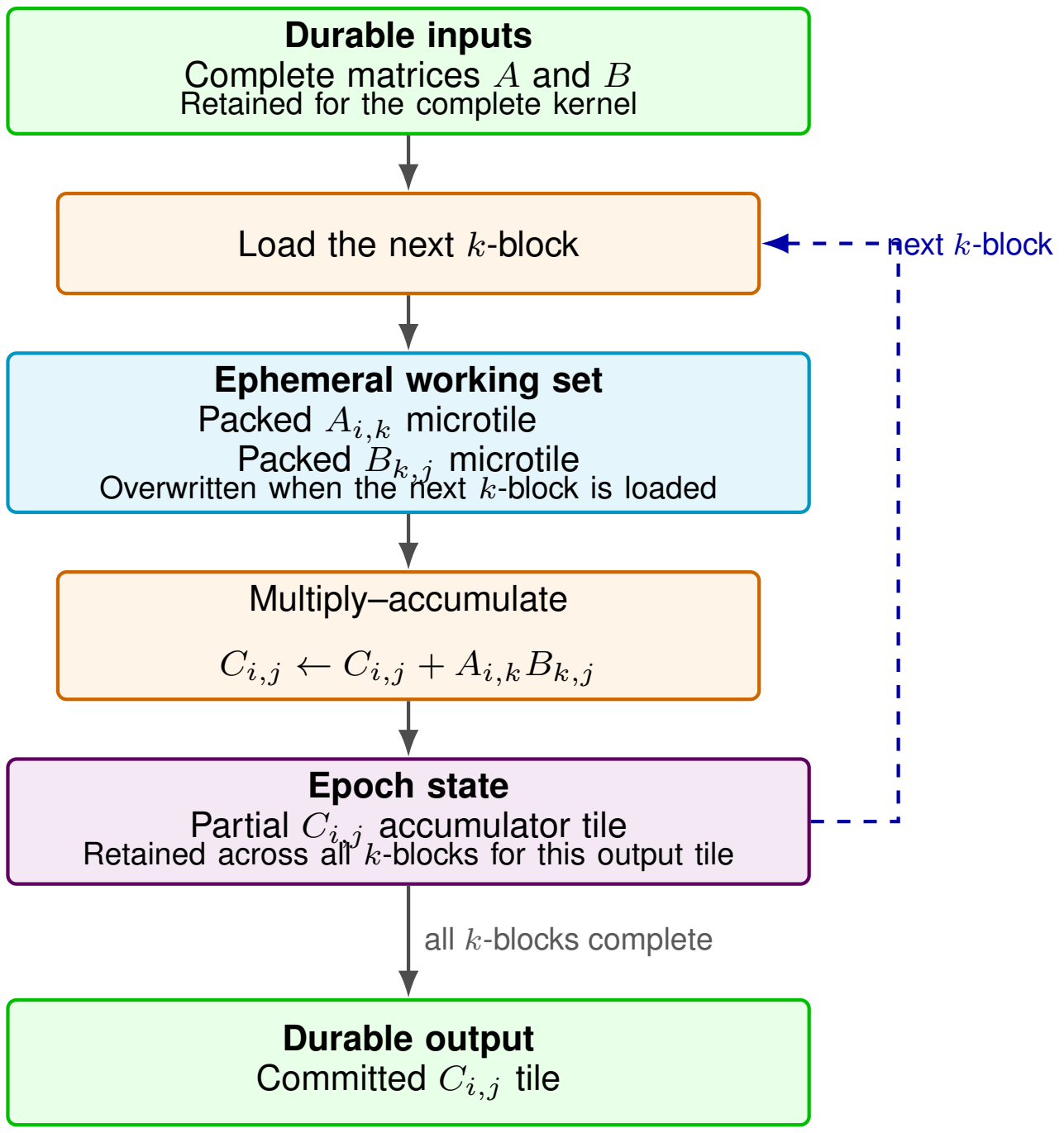

A tiled implementation, as outlined in Figure~\ref{fig:tiled-dataflow}, would change the lifetime requirements of these data structures. It would still keep the full $A$ and $B$ matrices, and the final $C$, in durable storage, however for each output tile, small sub $A$ and $B$ microtiles can be created in ephemeral storage and partial $C$ tiles can be accumulated in epoch class storage. In this configuration the microtiles are overwritten at the next inner step and the accumulator data will be committed after the output tile calculation has completed. The same computation has therefore acquired explicit phase boundaries and small working sets with bounded lifetimes.

Our prototype approach records each allocation, range access, compute interval, phase, barrier, hint, and free operation in a policy-independent trace. A lifetime analyser tracks each written cache-line version from its defining write to its last read or overwrite. The trace can then be replayed under multiple placement policies and device profiles without rerunning the application. This allows us to separate workload characteristics from policy decisions on where data structures are stored, allowing analysis of the potential performance, correctness, and energy impact of temporal memory approaches.

This approach can be generalised. For instance, in an AI context, attention kernels can separate persistent weights, step-scoped key/value state, and short-lived score tiles. Graph analytics can
separate a durable graph, epoch-scoped visited state, and ephemeral frontiers. Hash joins can separate durable inputs, an epoch hash table, and ephemeral match buffers. The opportunity is greatest where a transformation both improves locality and makes data expiration points explicit.

\subsection{Example case studies}

The lifetime abstraction becomes useful only if it maps onto recognisable execution boundaries. Table~\ref{tab:case-studies} gives three candidate mappings. The appropriate placement of concrete implementations would depend on actual device characteristics, workload, and costs of moving data between classes for the memory technology being used.

\begin{table*}[t]
\caption{Candidate lifetime contracts in example workloads.}
\label{tab:case-studies}
\centering
\footnotesize
\begin{tabularx}{\textwidth}{@{}p{0.13\textwidth}YYYY@{}}
\toprule
Workload &
Durable data &
Epoch data &
Ephemeral data &
Semantic boundary \\
\midrule

Tiled attention & Weights, input tensors, and committed outputs & Running softmax statistics and partial-output accumulator & Score/probability block and temporary reductions & A scratch tile becomes dead after its contribution has been incorporated into the running softmax and output \cite{dao2022} \\

\addlinespace

LLM serving &
Model weights, recovery state, and reusable prefix data &
Key/value pages belonging to active requests &
Rejected speculative branches and temporary scheduling state &
Request completion, branch rejection, or promotion into a shared prefix cache
\cite{kwon2023} \\

\addlinespace

Training &
Parameters, optimiser state, and committed checkpoints &
Activations, gradient buckets, and step-scoped state &
Fused-kernel scratch and temporary reductions &
Last backward consumer, completion of a collective, or successful step commit \\

\bottomrule
\end{tabularx}
\end{table*}

Tiled attention provides a particularly direct extension of the matrix-multiplication case study. IO-aware attention forms score or probability blocks, uses them to update running normalisation statistics and a partial output accumulator, and then discards them rather than materialising the complete attention matrix\cite{dao2022}. A lifetime contract can place each score block and its temporary reductions in ephemeral storage until that update completes. The running normalisation state and partial output instead require epoch retention until all key/value blocks for the query tile have been processed.

Key/value caches illustrate a more dynamic case. Pages belonging to an active request may be allocated to an epoch class, while rejected speculative branches can be invalidated immediately. If a request prefix is selected for reuse, its pages can be promoted to a stronger class. Page-granular systems such as PagedAttention already expose allocation, sharing, and eviction events that a retention-aware runtime could use \cite{kwon2023}.

These examples must ultimately be tested using end-to-end traces. Relevant measurements include the fraction of writes served by each class, worst-case rather than average object lifetime, promotion and refresh traffic, capacity pressure, application throughput, and tail latency.

\section{A Prototype That Separates Evidence from Assumption}

We have built a prototype stack with three parts. The first part is a C ABI and C++ interface that provides programming functionality to define data regions, buffers, specify semantic hints, mapping, refreshes, promotions, and collect statistics on data structure temporal characteristics. We have also developed an event recorder and replay simulator model that captures cache line-level retention, data class capacity, latency, refresh, migration, and failure events that would happen on expired data reads. To complement this we have also developed a prototype FPGA implementation which maps the three logical retention classes onto separate high bandwidth memory (HBM) pseudochannels and uses an FPGA copy kernel to exercise promotion paths. Reclassification allocates space in the destination class and invokes a 512-bit FPGA copy kernel. Thus, this experiment measures the practical cost of promotion and demotion through the proposed software and controller interface.

The reported experiments exercise trace recording and replay, together with migration between logical retention classes implemented using HBM pseudochannels. The FPGA measurements characterise allocation and reclassification costs, while the replay results explore retention safety and energy under an illustrative device profile. These results support the software interface and evaluation approach. They do not yet establish end-to-end enforcement of event-based validity under execution overruns, concurrent consumers, or changing operating conditions.

Table~\ref{tab:u280-migration} reports bidirectional transfers among all three logical classes. Each entry gives the range across the six directed class pairs. For each direction and size, five warm-up transfers were followed by 50 measured transfers. Measurements were collected on an AMD Alveo U280 using XRT~2.14.0 and the  \texttt{xilinx\_u280\_gen3x16\_xdma\_1\_202211\_1} platform, generated with Vivado~2022.1.1. The HLS migration kernel used the platform's default 300~MHz compute-unit clock, while the HBM subsystem operated at 450~MHz. Copy-path is the host-observed FPGA kernel submission and completion interval used to calculate the observed bandwidth. Application-visible additionally includes destination allocation and runtime reclassification overhead so represents the bandwidth the application observes.

\begin{table}[t]
\caption{Measured U280 migration throughput in GB/s. Ranges span the median
result for each of the six directed class pairs.}
\label{tab:u280-migration}
\centering
\scriptsize
\setlength{\tabcolsep}{4pt}
\begin{tabular}{rcc}
\hline
\textbf{Transfer size} &
\textbf{Copy-path throughput} &
\textbf{Application-visible throughput} \\
\hline
256 KiB & 3.17--3.36 & 1.98--2.89 \\
1 MiB & 4.33--4.40 & 2.64--4.06 \\
16 MiB & 4.79--4.87 & 3.01--3.10 \\
64 MiB & 4.86--4.94 & 3.25--3.28 \\
128 MiB & 4.88--4.96 & 3.26--3.33 \\
\hline
\end{tabular}
\end{table}

For large transfers, the copy path approaches 4.9~GB/s and shows little dependence on migration direction. End-to-end throughput approaches 3.3~GB/s because destination allocation and runtime operations remain on the critical path. For a 128~MiB transfer, destination allocation takes approximately 11~ms, the kernel interval takes 27.0--27.5~ms, and the complete reclassification takes 40.3--41.2~ms. We see that the fixed costs have a larger effect on small transfers. These measurements indicate that promotion is feasible but should normally operate on sufficiently large regions, reuse preallocated destination storage, or overlap migration with computation.

Table~\ref{tab:matmul} reports a complementary simulation experiment in which a \(32\times32\) matrix multiplication was traced using a panel width of four and three output-tile sizes. The simulator selected the shortest-retention classes that safely covered the observed temporary-data lifetimes. For each tile size, the sweep retained the capacities, channel counts, latencies, and energy parameters of the equal-resource profile while varying only the ephemeral and epoch retention windows. The table reports conservative safe windows selected from these profile variants.

The replay used an illustrative equal-capacity, equal-channel profile. Each retention class was provisioned with 1,048,576 64-byte lines (64 MiB) and one independent access channel. With a modelled cycle time of 1 ns, the ephemeral class had a retention window of 128 cycles, two-cycle reads and writes, and respective read and write energies of 1.0 and 2.0 pJ per 64-byte line. The epoch class had a retention window of 4,096 cycles, read and write latencies of three and four cycles, and respective energies of 1.4 and 4.5 pJ. The durable class had unbounded modelled
retention, four-cycle reads, 12-cycle writes, and respective energies of 2.0 and 12.0 pJ.

Each modelled operation additionally incurred 0.20 pJ of controller energy and 0.05 pJ of metadata energy. ECC added 0.10 pJ to a read and 0.20 pJ to a write. Consequently, excluding static energy and migration, the total dynamic charges for ephemeral, epoch, and durable reads were 1.35, 1.75, and 2.35 pJ, respectively; the corresponding write charges were 2.45, 4.95, and 12.45 pJ. Migration adds one setup cycle and 1.0 pJ in addition to the source-read and destination-write costs.

The profile assigned static powers of 0.001, 0.002, and 0.003~mW to the ephemeral, epoch, and durable classes, respectively. The totals in
Table~\ref{tab:matmul} include both dynamic and static modelled energy although these particular replays performed no migrations.

These parameters are reproducible sensitivity-analysis assumptions rather than measurements of the U280 or a fabricated STT-MRAM device. The results should therefore be interpreted as a design-space demonstration: if a future device provides comparable class trade-offs, the lifetime contracts expose sufficient temporal structure to exploit them.

\begin{table*}[t]
\caption{Preliminary deterministic replay data for a tiled matrix multiplication.
Reductions are relative to all-durable placement and use an illustrative device profile.}
\label{tab:matmul}
\centering
\footnotesize
\begin{tabular}{rrrrrrrr}
\toprule
Tile & Micro max & Accum.\ max & Retention & Input reads&
Cycles  & Energy & Reduction \\
 & lifetime (cycles) & lifetime (cycles) & (cycles; eph./epoch)) & (bytes) &
mix/dur.  & mix/dur. (pJ) & cycles / energy \\
\midrule
4 & 1,130 & 960 & 2,048 / 4,096 & 131,072 &
401,408 / 779,264 & 271,925 / 578,218 & 48.5\% / 53.0\% \\
8 & 4,074 & 4,160 & 4,096 / 8,192 & 65,536 &
352,256 / 648,192 & 232,309 / 456,190 & 45.7\% / 49.1\% \\
16 & 15,338 & 15,936 & 16,384 / 32,768 & 32,768 &
327,680 / 582,656 & 212,500 / 395,176 & 43.8\% / 46.2\% \\
\bottomrule
\end{tabular}
\end{table*}

The experiment exposes a useful systems trade-off. Larger tiles halve durable input traffic as the tile grows from four to eight to 16, and reduce modelled cycles and energy required for the activity. They also require larger temporary capacity and longer retention. Smaller tiles fit more aggressive memory classes but give up reuse. Retention, capacity, and locality are consequently coupled optimisation variables rather than independent design choices.

Our evidence has four explicit sets of boundaries that define the scope of the findings:

\begin{itemize}
\item \textbf{Measured on the FPGA:} allocation, migration, and application-visible reclassification timing.
\item \textbf{Observed from benchmark execution:} access ordering, byte ranges, phase boundaries, barriers, and data-structure identities.
\item \textbf{Derived or modelled:} cache-line-version lifetimes under the conservative timing model, retention safety, class access latency, energy, capacity contention, refresh, and migration energy.
\item \textbf{Not yet validated:} fabricated low-retention STT-MRAM arrays, temperature-dependent tail failures, and full controller energy.
\end{itemize}

This separation is deliberate. A simulator can reject unsafe contracts and quantify break-even conditions before fabrication, but it cannot substitute for device characterisation.

\section{Costs and Failure Modes}

Lifetime contracts introduce architectural and software costs that could eliminate their expected benefit. A practical system must therefore retain an all-durable compatibility path and make the overhead of each additional mechanism visible.

Physically differentiated retention classes require multiple homogeneous arrays or banks, potentially with different cell designs, peripheral circuits, timing parameters, error-correction requirements, and characterisation procedures. This increases design and verification complexity and may reduce the capacity available to any one class. Restricting the system to a small number of well-characterised classes - for example ephemeral, epoch, and durable - is more realistic than exposing many fine-grained retention values.

Furthermore, the controller must record where data are placed and when their guarantees approach expiration. Per-cache-line deadlines would introduce significant metadata and lookup overhead. Region-level metadata and deadline buckets can reduce this cost, although coarse tracking may cause unnecessary refreshes or promotions. Migration must also preserve virtual addresses so that application pointers and accelerator descriptors do not become invalid when the physical bank changes.

Finally, refresh and promotion consume memory bandwidth, energy, and controller queue capacity. Promotion can be especially expensive because it requires allocation in the destination class followed by a complete data copy. The measured U280 migration curve can provide an initial cost model, but it does not establish that promotion is beneficial. A policy is worthwhile only when energy savings match:

\begin{equation}
E_{\mathrm{saved\ writes}} >
E_{\mathrm{refresh}} +
E_{\mathrm{migration}} +
E_{\mathrm{metadata}} +
E_{\mathrm{error\ handling}}.
\label{eq:energy-break-even}
\end{equation}

An analogous condition must hold for execution time and foreground memory bandwidth. A nominally faster short-retention bank can still reduce system performance if it creates bank hot spots, additional queueing, or frequent migration.

Execution time is also uncertain. Pre-emption, contention, data-dependent control flow, and changes in input size can cause an object to outlive its profiled lifetime. A \texttt{required} contract must remain correct under these conditions. The runtime must use conservative margins, monitor approaching deadlines, and refresh or promote data before validity can be lost. A \texttt{preferred} contract may instead fall back to a stronger class when the prediction is uncertain.

Concurrency creates a related problem. A phase may appear complete while another thread, accelerator stream, DMA operation, or tenant still holds a reference to its data. Phase completion must therefore be connected to synchronisation, coherence, and ownership transfer. The controller must not invalidate a region until all required consumers have retired.

Finally, adoption cannot depend on every application being rewritten immediately. Unannotated allocations should default to durable memory. Programmer annotations, profiling, and compiler analysis can then introduce more aggressive placement incrementally. Incorrect hints may reduce efficiency, but they must not compromise correctness.

Expiration must also not be confused with secure deletion. A contract states when data no longer require correctness, but it does not guarantee that residual contents are unreadable. Confidentiality still requires explicit scrubbing, encryption, or cryptographic erasure.

These costs establish the appropriate comparison for evaluation: all-durable placement, a static mixed policy, an oracle bound, and a practical online policy with metadata, capacity limits, refreshes, migrations, queueing, and fallback behaviour enabled.

\section{Correctness Comes Before Efficiency}

A retention class is a statistical guarantee, not a timer after which every bit flips. Memory retention in devices such as STT-MRAM varies with temperature, process, aging, data pattern, and cell history. Read disturb phenomenon and write failures further complicate this error surface. Thermal-aware and adaptive error-correction techniques illustrate why reliability mechanisms must be part of the architecture rather than an afterthought~\cite{wu2019}.

The controller should therefore expose a guaranteed interval at a target uncorrectable-error probability, including guard bands, to provide programmers with architectural commitments. Metadata can be managed at cache-line or region granularity, with deadline buckets avoiding a full-memory scan. Before a deadline, the policy associated with the memory will choose refresh, promotion, replication, or recomputation. Capacity pressure must trigger explicit behaviours: silently placing an object in a weaker class because its requested bank is full would violate the programmer contract.

Our present stochastic simulator is useful for testing control flow but is not yet a physical reliability model. It is configured with an exponential survival curve that treats projected data retention as a characteristic mean; at one such interval the raw failure probability is already $1-e^{-1}\approx63.2\%$. That convention is unsuitable for reporting realistic memory reliability without calibration, ECC, and tail-distribution measurements from physical devices. Until measured devices are available, we use deterministic guaranteed windows for correctness studies and label energy results as sensitivity analyses.

Our programming approach also introduces a semantic failure mode. A programmer may underestimate a lifetime, or an input may make a phase run longer than profiling predicted. Required contracts must remain correct despite such a programming mistake. The runtime can choose a stronger class initially, monitor elapsed time, or promote data before the conservative deadline. Hints may improve efficiency, but correctness cannot depend on perfect prediction.

\section{Temporal Memory as a Transformation}

Multiple retention banks alone are an architectural option or approach. A lifetime-aware cache alone is a policy approach. The transformative opportunity is the combination of these with a stable programming contract and algorithms that deliberately expose temporal structure. This can make the physical property of retention usable without forcing application developers to understand magnetic energy barriers or temperature margins.

Such an abstraction could also outlive any one technology. Retention-aware regions can target STT-MRAM today, other emerging memories tomorrow, or even conventional systems in which the contract governs refresh, replication, power-gating, and recomputation. The overarching idea is that data correctness has a duration, and that duration should influence resource allocation.

\section{Conclusion}

Memory systems expose address, capacity, and bandwidth, but largely conceal retention characteristics. Workloads make that omission costly because their data spans a wide spectrum of useful lifetimes. We propose treating lifetime as a first-class contract, implemented through homogeneous retention-class arrays and correctness-preserving control.

The prototype provides a region-oriented software interface, a trace-based framework for evaluating placement policies, and measured migration costs between logical retention classes. Tiled matrix multiplication illustrates how an established transformation exposes distinct temporary-data lifetimes. The modelled reductions in execution time and memory energy remain conditional projections pending calibration against fabricated devices. These results identify initial device targets and system costs for evaluating temporal memory. However, automatic enforcement of semantic validity under execution overruns and concurrency remains to be demonstrated.

\section*{AI Disclosure}
LLMs from the ELM service at University of Edinburgh were used to improve the quality of the text, as well as to generate the initial templates for the graphics included in this paper.

\section*{Acknowledgment}
The FPGA results in this article were collected on the ExCALIBUR H\&ES FPGA testbed\footnote{\url{https://fpga.epcc.ed.ac.uk/}}. The source code for the developed simulator and FPGA implementation is available here: \url{https://github.com/adrianjhpc/ProgrammableRetentionTiles}, release 0.1, git commit \texttt{fc445a6}. 

\bibliographystyle{IEEEtran}
\bibliography{references}

\begin{IEEEbiography}[\raisebox{40pt}{{\includegraphics[width=1in]{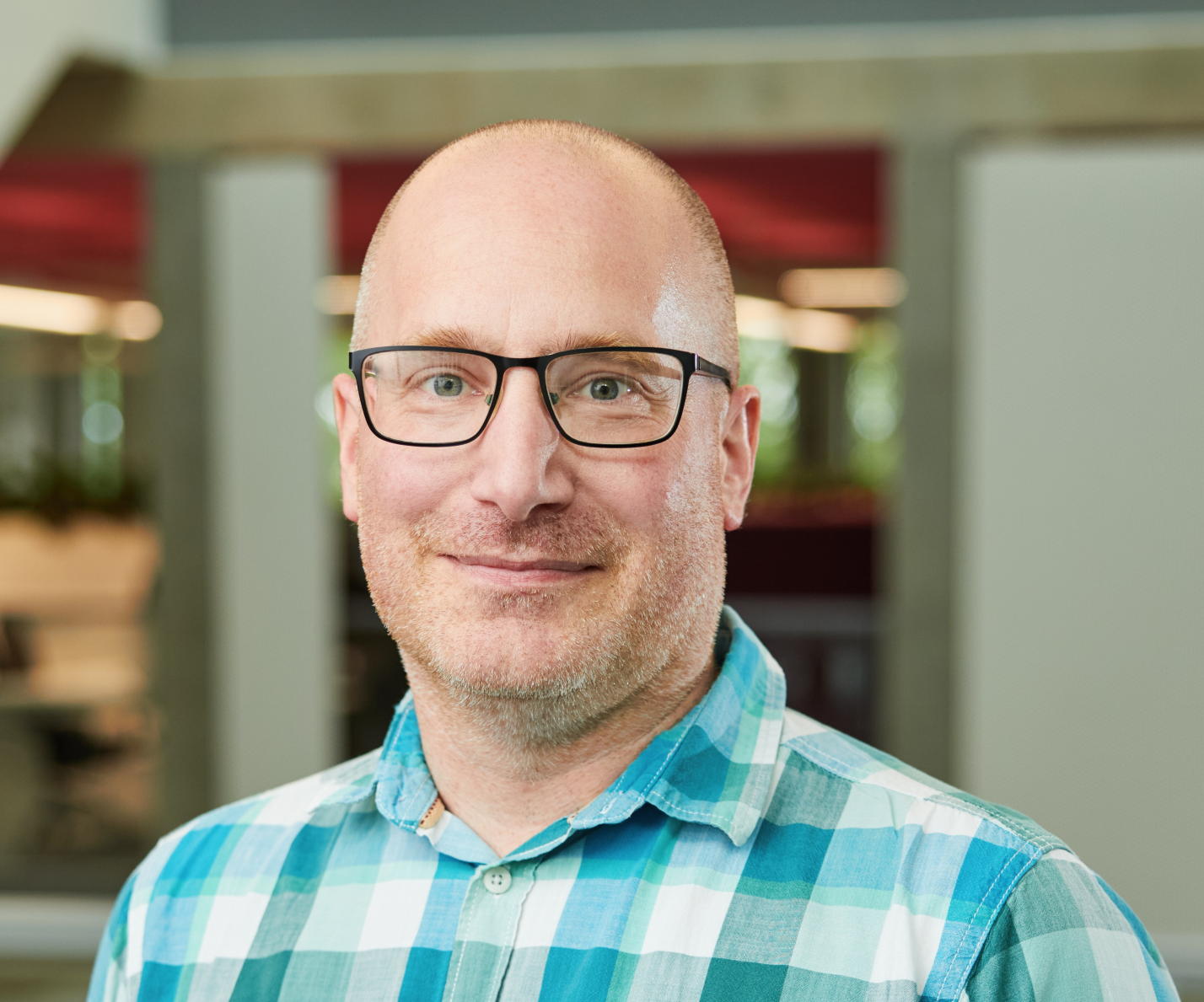}}}]{Adrian Jackson}
is Professor of High Performance Computing Technologies at EPCC, The University of Edinburgh. His research interests include computer architecture, memory systems, and large scale adoption of efficient computing technologies.
\end{IEEEbiography}

\end{document}